\documentclass{revtex4}
\usepackage{graphicx}

\usepackage{dsfont}
\usepackage{dcolumn}
\usepackage{bm}
\usepackage{amsmath}
\usepackage[T1]{fontenc}
\usepackage{xspace}

\usepackage{color}

\usepackage{ulem}

\newcommand{\ket}[1]
{
|#1 \rangle
}

\newcommand{\expect}[1]
{
\langle #1  \rangle
}

\begin{document}

\title{Spin polarisation with electron Bessel beams
}
\author{P. Schattschneider}
\affiliation{Institut f\"ur Festk\"orperphysik, Technische
Universit\"at
Wien, A-1040 WIEN, Austria}
\affiliation{USTEM, Technische Universit\"at Wien, A-1040 WIEN, Austria}
\author{V. Grillo}
\affiliation{CNR-Istituto Nanoscienze, Centro S3, Via G Campi 213/a, I-41125 Modena, Italy}
\affiliation{CNR-IMEM, Parco delle Scienze 37a, I-43100 Parma, Italy}

\begin{abstract}
The theoretical possibility to use  an  electron microscope as a spin polarizer is studied.  It turns out that a  Bessel beam  passing a standard magnetic objective lens is intrinsically spin polarized. In the limit of infinitely small detectors on axis, the spin polarisation tends to 100 \% . Increasing the detector size, the polarisation decreases rapidly, dropping below $10^{-4}$ for  standard settings of medium voltage microscopes. For extremely low voltages, the figure of merit increases by two orders of magnitude, approaching  that of existing Mott detectors. Our findings may lead to new desings of spin filters, an attractive option in view of its inherent combination with the electron microscope, especially at low voltage.
\end{abstract}


\maketitle 

\section*{Introduction}
After the prediction\cite{Bliokh2007} and the experimental demonstration\cite{UchidaNature2010,VerbeeckNature2010} of free electrons carrying orbital angular momentum (OAM),  these vortex beams---as they are called---are increasingly attracting interest.  They are characterized by a spiralling wavefront, similar to optical vortices\cite{NyeBerry1974}. In parallel to their OAM, they carry  magnetic moment that is  independent of their  spin polarization. 
Their  potential  ranges from the study of Landau states in an interaction-free environment\cite{SchattschneiderNatComm2014}, over probing chiral specimens with elastic or inelastic scattering on the nanoscale, to manipulation of nanoparticles, clusters and molecules, exploiting the magnetic interaction and the transfer of angular momentum\cite{VerbeeckAdvMat2013}. Bliokh et al.\cite{BliokhPRL2011} discovered an unexpected intrinsic spin-orbit coupling (SOC) in relativistic vortex electrons, and proposed to use this effect in a spin filter.

Despite the  statement of Bohr and Pauli that Stern-Gerlach based spin separation for electrons cannot work\cite{Pauli1964},   it has been argued that spin separation or filtering of electrons is possible in particular  geometries\cite{BatelaanPRL1997,GarrawayPRA1999}. The argument has been debated, see e.g.\cite{RutherfordJPhys1998}, and it seems that the effect exists but is too small to be exploited with present day technology. In later papers, interesting alternatives relying on inhomogeneous magnetic fields with cylindrical symmetry\cite{GallupPRB2001,McGregorNJP2012}, or on SOC in an electron transparent medium\cite{ChernPRL2010} were discussed. But as of now, no Stern-Gerlach design of a  spin polarizer for free electrons was successful.  

In an elegant derivation Karimi et al.~\cite{Karimi2012} have shown that crossed electric and  magnetic quadrupole fields correspond to an optical q-plate with $q=-1$. In combination with electron vortex beams, this opens the possibility to couple the spin of free electrons to the spatial degree of freedom, and so design a spin filter~\cite{GrilloNJP2013}.

The proposal is an analogue to the spin-to-orbital momentum conversion (STOC) of optical beams\cite{MarrucciPRL2006,GrilloUM2014}. 
The essential driving agent for the STOC process is a non-vanishing Berry connection\cite{Bliokh2007}. This translates either into the connection $\mathcal A(\bf p)$ in momentum space, or in the presence of a magnetic field into the vector potential $\bf A(r)$ in real space.
 $\bf A(r)$ with cylindrical symmetry over the propagation axis is equivalent to an optical phase plate with $q=1$.  Such a field can be used as a STOC device quite similar to the optics case because  the total angular momentum $L + S$  is a constant of motion\footnote{This is the angular momentum in the corotating Larmor frame. The fact that the {\it mechanical} total angular momentum $\mathcal L+S$ in the laboratory frame is not conserved has no consequence on the STOC process.} . This configuration is encountered in a magnetic round lens used in electron microscopes. Thus, it seems that  electron microscopes are intrinsic spin polarizer for electron vortex beams.
 
 Instead of solving the Dirac equation in the inhomogeneous lens field, which is a complicated numerical problem. Here, we aim simply at an estimate of the order of magnitude of the STOC effect. For an intuitive understanding of the underlying principle, we shall  make use of the concept of Bohmian trajectories [Bliokh].



\section*{Results}
\begin{figure}[htbp]
	\centering
		\includegraphics[width= 0.55 \columnwidth]{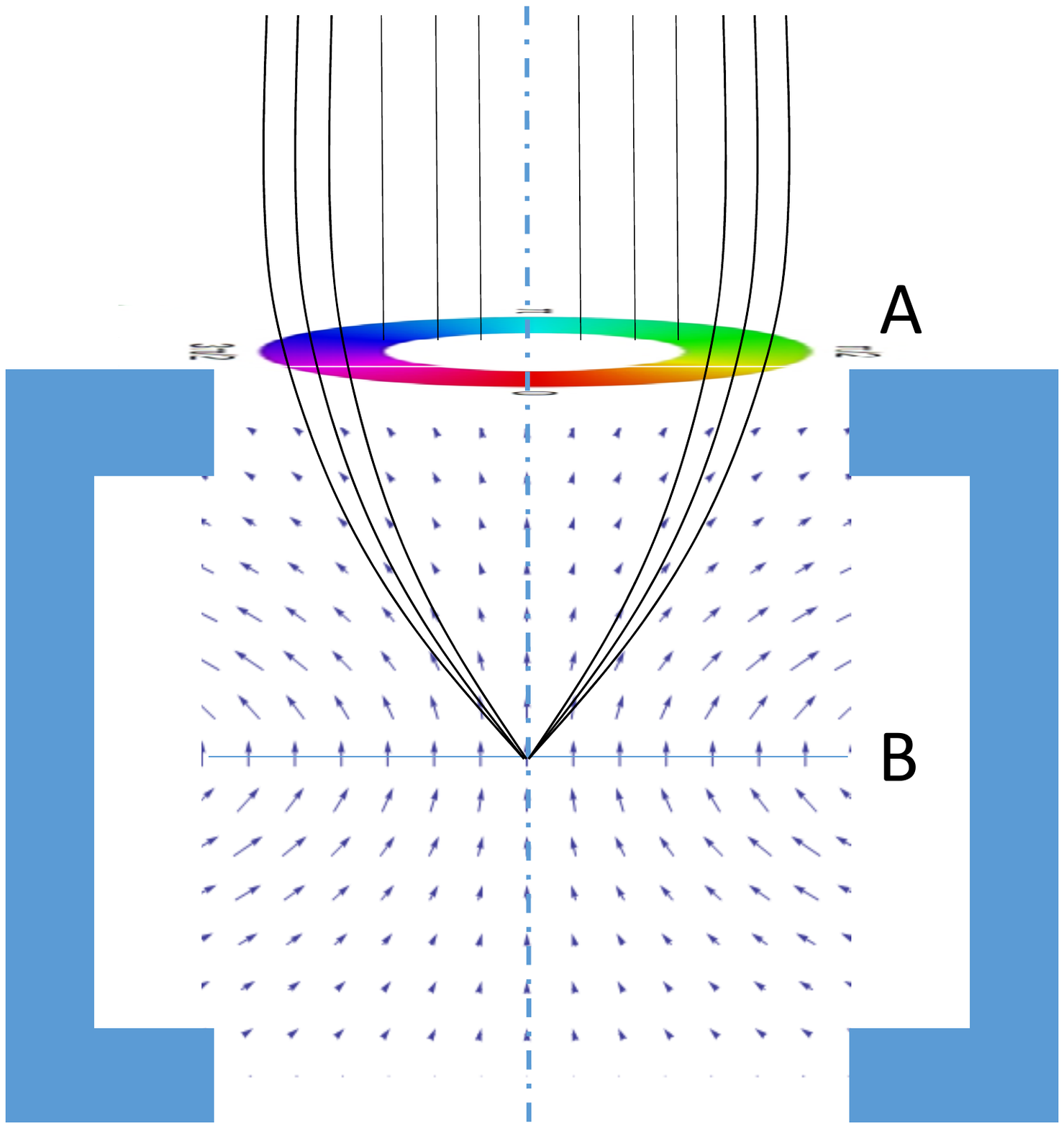}
		\includegraphics[width=0.4 \columnwidth]{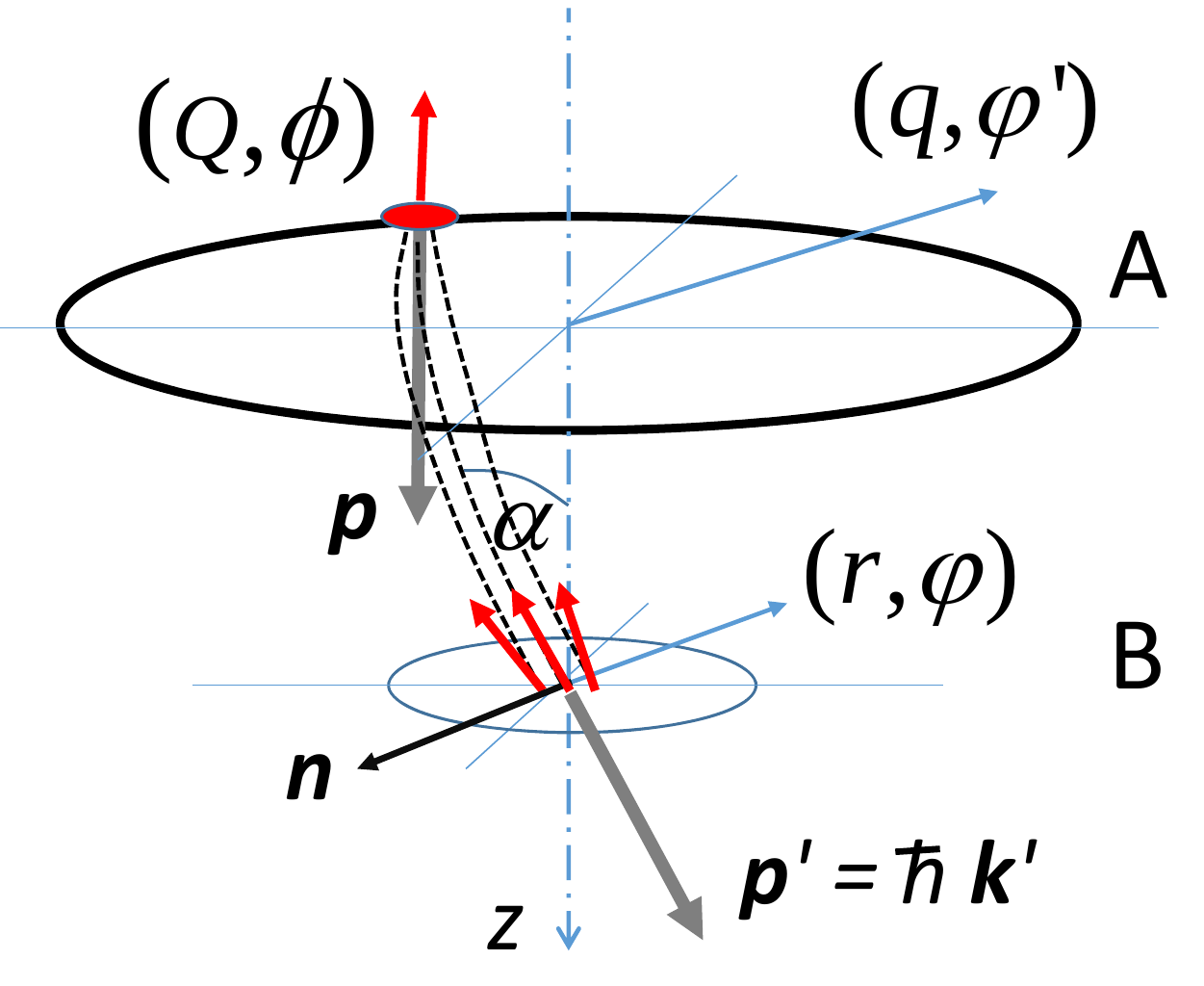}
	\caption{Schematic of the proposed geometry. Left: A plane wave incident in the front focal plane of a magnetic round lens is limited by an annular aperture selecting a narrow range of lateral momenta.  A spiral phase plate (or a similar vortex generating device) imprints an azimuthal phase ramp $\phi$ thus creating a  vortex beam with OAM=1. The phase is indicated by the rainbow colors along the ring aperture. The  image plane contains the Fourier transform,  a Bessel beam $J_1$. The yoke and the magnetic field are indicated. Right: Cylindical coordinates in planes A and B (blue arrows). The pointlike source (red dot) with coordinates $Q,\phi$ on the ring aperture (the integrand in Eq.~\ref{eq:psiQ}) creates a plane wave with wave vector ${\bf k}'$ in plane B. In the classical description, the momentum $\bf p$ of the  particle in A is tilted by the Lorentz force of the magnetic lens field to ${\bf p}'= \hbar {\bf k}'$ at B (grey arrows). The spin vector of a hypothetical spin polarized electron (red arrows) performs a precession in the magnetic field when the outgoing wave propagates to plane B, conserving  helicity. Three Bohmian trajectories are shown (dashed lines) together with the $r$-dependent spins, slightly deviating from the tilt angle $\alpha$ over the unit vector $\bf n$. Vectors are not to scale.}
	\label{fig:Sketch}
\end{figure}

We assume a  a plane wave parallel to the optic axis, $w \, e^{i k z}$, incident in the front focal plane A of a round magnetic lens, as sketched in Fig.~\ref{fig:Sketch}. $w=(a_+,a_-)^T$ is the Pauli spinor in the eigenbasis of the Pauli matrix $\sigma_z$ ,
and the quantisation axis $z$ is the optic axis (but any other basis would also work). 
The coefficients $a_\pm$ are normalised as $|a_+|^2+|a_-|^2=1$.  
The spinor is assumed to be independent of the spatial coordinates of the plane wave, i.e. the plane wave has unique spin polarisation.

The wave function is prepared by a ring aperture imprinting an azimuthal phase ramp of $2 \pi$. This can be achieved by a spiral phase plate, a holographic mask, or any other method suitable to produce beams with topological charge. For the moment, we assume an infinitesimally narrow ring  of radius $Q$ that prepares the electron in plane A located at position $z=0$. 

Then  the wave function after passing the ring aperture is
$$
\ket \Psi=(a_+ \ket \uparrow + a_- \ket \downarrow )\otimes \ket {\psi_A}
$$
with the spatial part given in cylindrical coordinates $(q, \varphi)$ as an azimuthal superposition of point sources on a ring with radius $Q$, 
\begin{equation}
\psi_A(q,\varphi)=\frac{1}{\sqrt{2 \pi Q}}
\int_0^{2\pi}  
 \delta(q-Q)\delta(\varphi-\phi)e^{i\phi} \, d\phi.
\label{eq:psiQ}
\end{equation}

 In the magnetic field of the lens, the wave propagates from plane A to  plane B.

The point $(Q,\varphi)$ can be interpreted as the source of a bundle of Bohmian trajectories that cross the optic axis close to plane B, with convergence angle $\alpha$. Since the angle between spin and momentum remains constant in the magnetic field, the spin direction in B depends on the position, as sketched in Fig.~\ref{fig:Sketch}. In the narrow radial range of a few nm centered at the optic axis that interests us later, the deviation of the spin directions from $\alpha$ is negligible, so we may consider $\alpha$ as independent of the radial distance $r$ in plane B.

Thus, on arrival at plane B the initial spin will have been rotated by  $\approx\alpha$ over some direction $\vec n$ on the Bloch sphere.
The operator for the spinor rotation   is
\begin{equation}
R_{{\vec n}}(\alpha)=e^{-i\frac{\alpha}{2} {\vec \sigma}{\vec n}},
\label{Rn}
\end{equation}
with the Pauli matrices $\vec \sigma$.

The transfer of the {\it spatial} part, i.e. a point source $\psi_{Q,\phi}$ from the front focal plane A to the object plane B can be described by a 2D Fourier transform. Combining this with the spinor rotation, the point source at $(Q,\phi)$ in plane A has evolved into 
\begin{eqnarray}
\tilde\psi_{Q,\phi}(r,\varphi)&=&R_{{\vec n}}(\alpha) \, w \int q\,dq\,d\varphi' \, e^{i {\bf q r}} \psi_{Q,\phi}(q,\varphi')= \nonumber \\
&=&R_{{\vec n}}(\alpha) \, w \, e^{i Q r \cos(\phi-\varphi)}e^{i \phi} e^{i k z} 
\label{psitildeQ}
\end{eqnarray}
 in plane B.
As expected, the spatial part is a plane wave with wave vector $ {\bf k}'=(Q \cos \Phi, Q \sin \Phi,k)$ and a rotated spinor~\footnote{In paraxial approximation. The azimuth $\varphi$ refers to the corotating Larmor system, that takes into account the image rotation in the lens~\cite{Glaser1952}.}.

The total wave function $\psi_B$ in plane B is the azimuthal integral over all partial plane waves Eq.~\ref{psitildeQ},
\begin{equation}
\psi_B(r,\varphi)=\tilde \psi_A= \frac{1}{2 \pi} \int_0^{2\pi} d\phi  {\tilde\psi_{Q,\phi}}(r,\varphi) := T \, w \, \psi_A 
\label{eq:psiB}
\end{equation}
The transfer matrix $T$ defined here is derived in the methods section: 
\begin{equation}
T=
 i\,  e^{i k z}\left( \begin{array}{cc}
 J_1(Qr)e^{i\varphi}(\cos \frac{\alpha}{2} -i \sin \frac{\alpha}{2} \cos \theta) & - J_0(Qr) \sin \frac{\alpha}{2} \sin \theta \, e^{-i\phi_0} \\
 J_2(Qr)e^{2i \varphi}\sin \frac{\alpha}{2} \sin \theta \, e^{i\phi_0} &  J_1(Qr)e^{i \varphi}(\cos \frac{\alpha}{2} + i \sin \frac{\alpha}{2} \cos \theta)
 \end{array} \right) \, .
\label{T}
\end{equation}

Ignoring the common phase factor $i \, e^{ikz}$, the special case of a spin up (spin down) state at plane A  evolves into a state 
\begin{subequations}
\label{PlaneB}
\begin{align}
  \label{eq:planeBa}
\ket {\uparrow}\otimes \psi_A \, \rightarrow \,T\, \ket {\uparrow}= J_1(Qr)e^{i \varphi}\cos {\frac{\alpha}{2}} \ket {\uparrow} + J_2(Qr)e^{2i \varphi} \sin {\frac{\alpha}{2}} e^{-i \phi_0} \ket {\downarrow}
\\
  \label{eq:planeBb}
\ket {\downarrow} \otimes \psi_A \, \rightarrow \,T\, \ket {\downarrow}= -  J_0(Qr)\sin {\frac{\alpha}{2}} e^{i \phi_0} \ket {\uparrow}+  J_1(Qr)e^{i \varphi}\cos {\frac{\alpha}{2}} \ket {\downarrow}
\end{align}
\end{subequations}
at plane B, since in this case the rotation axis for the spinor is perpendicular to the $z$ axis, and so $\theta=\pi/2$. 

It is important to notice that the spin flip has introduced components with azimuthal phase ramps $4 \pi$ and 0, respectively in the two cases. Application of the operator for the orbital angular momentum, $- i \hbar \partial_\varphi$ shows that the OAM differs from $\hbar$, the OAM imprinted by the vortex mask. This is the consequence of the spin flip, or in other words the spin-to-orbit conversion induced by the magnetic field.

It is straightforward to calculate the expectation values of the OAM and the spin moment of the electron in plane B. For the spin up case, Eq.~\ref{eq:planeBa}, this is
$$
\expect {L_z}=\hbar (\cos^2 \frac{\alpha}{2} +2 \sin^2\frac{\alpha}{2})
$$
$$
\expect{ S_z} =\frac{\hbar}{2} (\cos^2 \frac{\alpha}{2} - \sin^2\frac{\alpha}{2})
$$
and 
$$
\expect {J_z} = \expect {L_z+S_z}=\frac{3 \hbar}{2}.
$$
For the spin down case, Eq.~\ref{eq:planeBb}
$$
\expect {L_z}=\hbar \cos^2 \frac{\alpha}{2} 
$$
$$
\expect{ S_z} =\frac{\hbar}{2} (\cos^2 \frac{\alpha}{2} + \sin^2\frac{\alpha}{2})
$$
and 
$$
\expect {J_z} = \expect {L_z+S_z}=\frac{\hbar}{2}.
$$
As dictated by the cylindrical symmetry of the problem, the $z$ component of the total angular momentum is constant.

The salient aspect of Eqs.~\ref{PlaneB} that allows spin filtering is the spatial separation of spin components in the Bessel functions $J_m$. 
We assume that the plane wave incident in plane A has no spin polarisation. This is justified when the source emits unpolarized electrons outside of the magnetic field of the lens. Unpolarized (or partially polarized beams) are described by a density matrix. In a $\ket{\uparrow}, \, \ket{\downarrow}$ basis the unpolarized wave in plane A has the density matrix
$$
\rho_A=\frac{1}{2}\psi_A \psi_A^*  \begin{pmatrix}
1& 0 \\
0 & 1
\end{pmatrix}.
$$

The density matrix in plane B is
$$
\rho_B :=\begin{pmatrix}
\rho_{\uparrow \uparrow} & \rho_{\uparrow \downarrow} \\
\rho_{\downarrow \uparrow}& \rho_{\downarrow \downarrow}
\end{pmatrix}= T \, \rho_A \, T^+.
$$
The diagonal matrix elements derive from Eq.~\ref{T},
\begin{subequations}
\label{rhopp}
\begin{align}
  \label{eq:rhoup}
\rho_{\uparrow \uparrow}=\frac{1}{2} \,(\cos^2 {\frac{\alpha}{2}}J_1^2(Qr) +\sin^2 {\frac{\alpha}{2}}J_0^2(Qr))
\\
  \label{eq:rhodown}
\rho_{\downarrow \downarrow}=\frac{1}{2} \,(\cos^2 {\frac{\alpha}{2}}J_1^2(Qr) +\sin^2 {\frac{\alpha}{2}}J_2^2(Qr))
\end{align}
\end{subequations}
The {\it differential longitudinal} spin polarisation 
in the object plane 
is given by the trace  of the Pauli matrix $\sigma_z$ over the spin variable $s$ when properly normalized:
\begin {eqnarray}
\label{polari}
\frac{\partial^2 p}{\partial{r}^2}&=&{\rm Tr}_s[\rho \, \sigma_z]/{\rm Tr}_s[\rho]  =\frac{\rho_{{\uparrow \uparrow}} - \rho_{{\downarrow \downarrow}}}{\rho_{{\uparrow \uparrow}} + \rho_{{\downarrow \downarrow}}} \\
&=&\frac{(J_0^2(Qr)-J_2^2(Qr))\sin^2 \frac{\alpha}{2}}
{(J_0^2(Qr)+J_2^2(Qr))\sin^2 \frac{\alpha}{2}+2 J_1^2(Qr)\cos^2 \frac{\alpha}{2}} \nonumber
\end{eqnarray}

\begin{figure}
	\centering
		\includegraphics[width=0.8 \columnwidth]{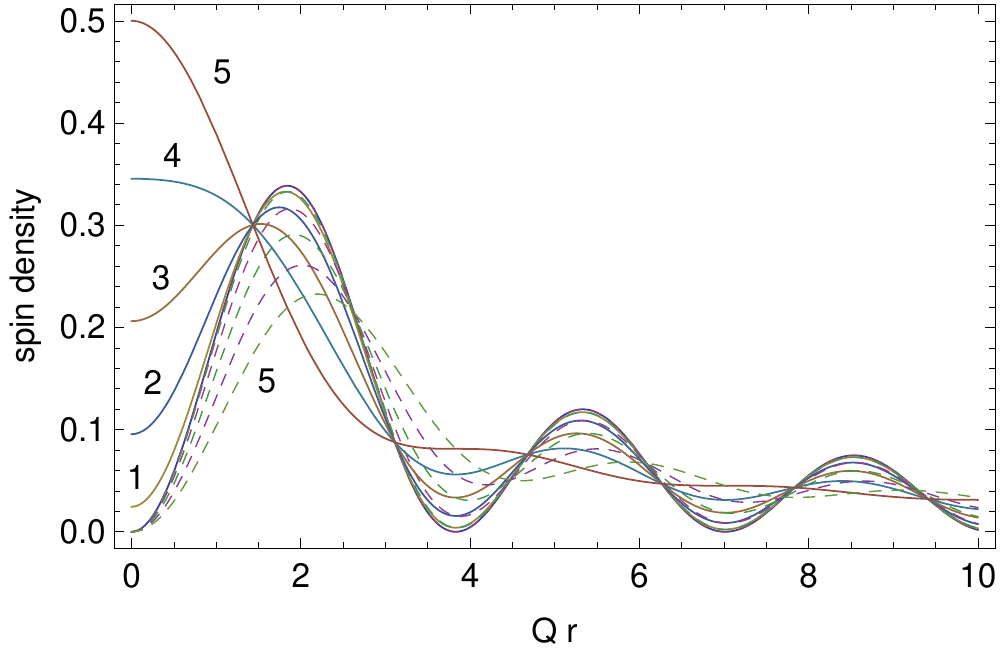}
		\vskip 0.5cm
		\includegraphics[width=0.8 \columnwidth]{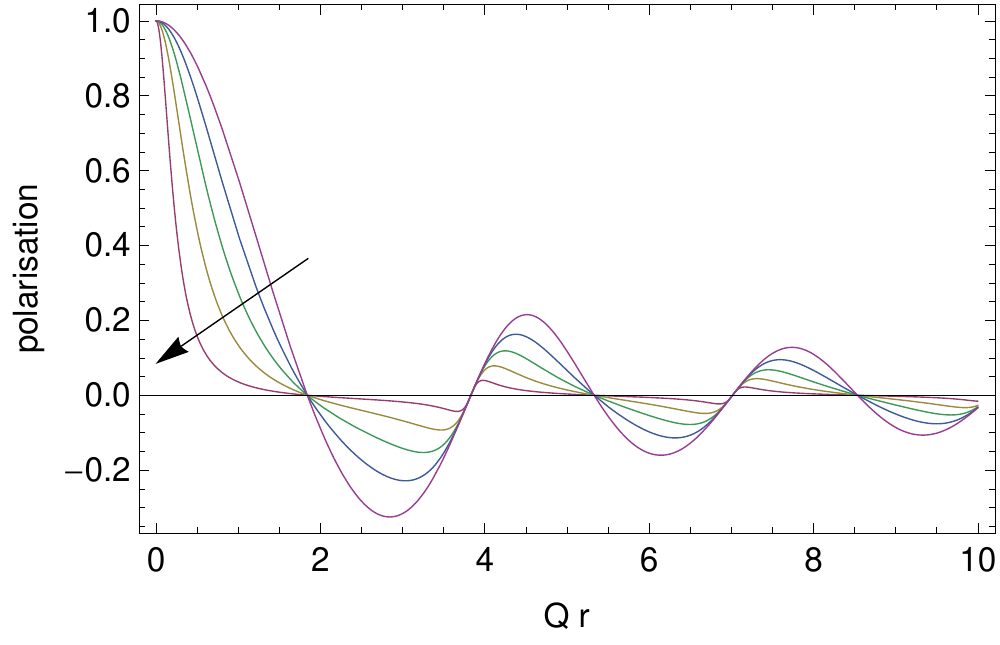}	
		\caption{a) Position dependent spin densities in the object plane, Eq.~\ref{rhopp}, as a function of the dimensionless variable $Q r$.  $\rho_{\uparrow \uparrow}$ full lines, $\rho_{\downarrow \downarrow}$ dashed lines for $\alpha= n \pi/10, n \in [0,5]$. Curves are labelled with $n$. b) differential spin polarisation, Eq.~\ref{polari}, again for $\alpha= n \pi/10, n \in [0,5]$ for  pointlike  detectors. At $r=0$ it is unity. $\alpha$ increases as indicated by the arrow.}
	\label{fig:polarisation}
\end{figure}

\begin{figure}
	\centering
		\includegraphics[width=0.8 \columnwidth]{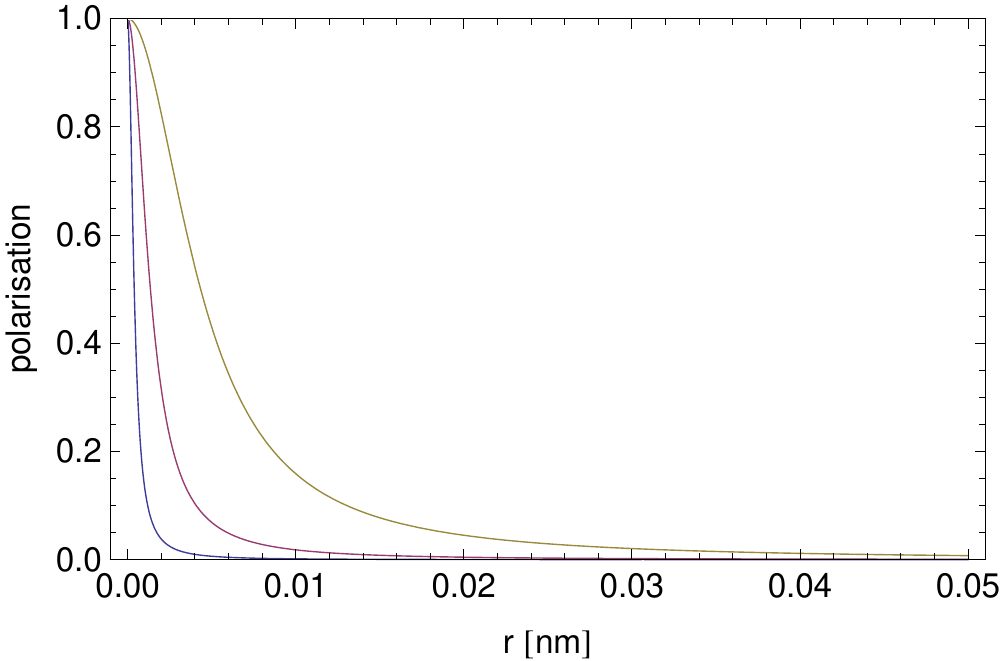}
		\caption{a) Polarisation as a function of detector radius $r$, Eq.~\ref{polari2}, for 200 (blue), 20 (magenta), and 2 kV (green) incident energy. The convergence angle is 50 mrad. 
		}
	\label{fig:polarisation2}
\end{figure}

\begin{figure}
	\centering
	\includegraphics[width=0.8 \columnwidth]{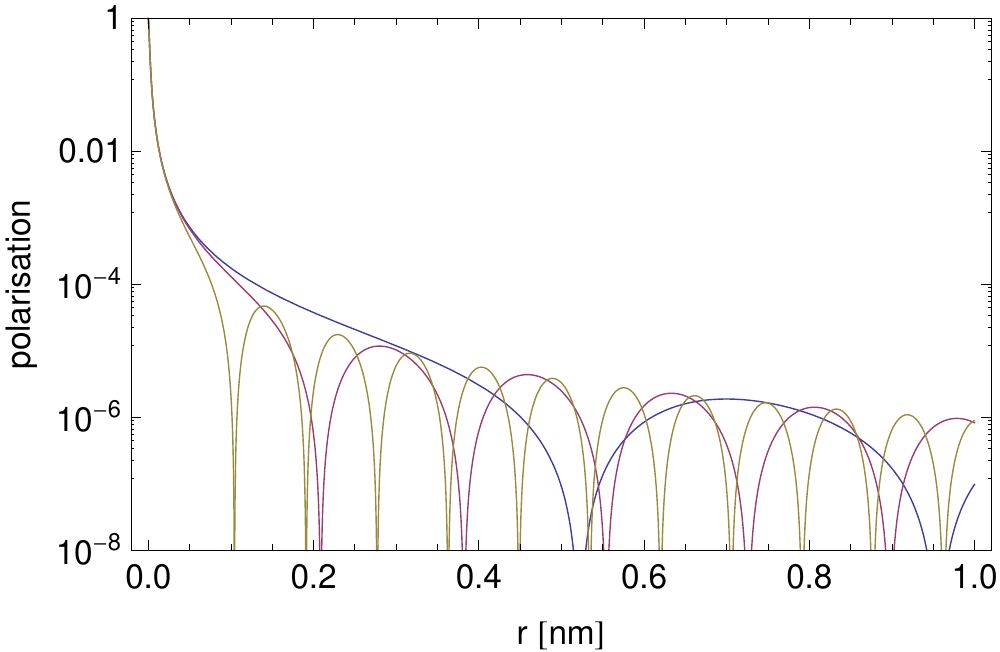}
		\caption{  Polarisation as a function of detector radius $r$, Eq.~\ref{polari2}, for 20 keV incident energy. The convergence angles are 10 (blue), 25 (magenta), and 50 mrad (green).}
	\label{fig:polarisation3}
\end{figure}
The spin densities Eq.~\ref{rhopp} are shown in Fig.~\ref{fig:polarisation}a for a variety of (quite high) convergence angles $\alpha$. (sandard values in conventional TEMs are $< 0.1$~rad). $\rho_{{\uparrow \uparrow}}$ and $\rho_{{\downarrow \downarrow}}$ converge to the same profiles for $\alpha \rightarrow 0$, as expected. 
Fig.~\ref{fig:polarisation}b shows the spin polarisation, Eq.~\ref{polari}. It oscillates and changes sign as a function of detector position. An infinitely narrow detector at the optic axis would serve as a perfect spin filter. 

An on-axis detector with finite area, say, of disk of radius $\Delta r$ integrates the signal over that area, giving a  measurable polarisation
\begin {equation}
p(\Delta{r})={\rm Tr}_{s,\Delta r}[\rho \, \sigma_z]/{\rm Tr}_{s,\Delta r}[\rho ].
\label{polari2}
\end{equation}
This is shown in Fig.~\ref{fig:polarisation2}. 
Lower energy electrons show stronger polarisation at the same convergence angle. The log scale shows clearer what can be expected. Even with very low energies of 2 keV, the polarisation is $\sim 10^{-3}$ at a detector radius of 0.1~nm and levels off to $10^{-5}$ at  1~nm.

The influence of the convergence angle is shown in Fig.~\ref{fig:polarisation3} for 20 keV. Interestingly, the tendency is counterintuitive: The smallest convergence angles show less oscillatory behaviour. A detector of 0.4~nm diameter would produce a spin polarisation of $10^{-4}$.

The dips in Fig.~\ref{fig:polarisation} are caused by the oscillations of the spin density. In a real experiment, they would be washed out by the deviation from pure Bessel functions, and by the limited coherence of the source.

\begin{figure}
	\centering
		\includegraphics[width=0.8 \columnwidth]{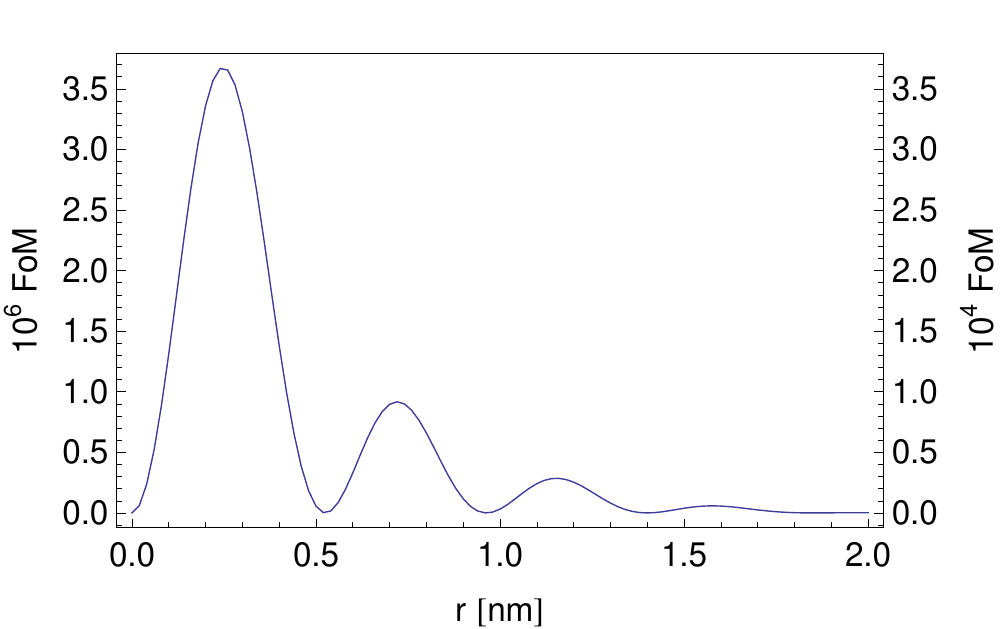}	
		\caption{Figure of Merit, Eq.~\ref{FoM}. Left ordinate: 20~kV incident energy and a ring aperture allowing convergence angles $\alpha$ from 8 to 12~mrad. Right ordinate: 200~V incident energy and a ring aperture allowing convergence angles $\alpha$ from 80 to 120~mrad. The low energy case has a FoM 100 times higher.}
	\label{fig:FoM}
\end{figure}

The detection efficiency is given by 
\begin{equation}
DE={\rm Tr}_{s,\Delta \bf r}(\rho)/I_0
\end{equation}
with the total intentsity $I_0$ in the input Bessel beam $J_1$.  Pure Bessel beams are unconfined and cannot be renormalized. However, experimental Bessel beams are always renormalizable because the ring apertures have finite width. 
 In that case, 
a figure of merit can be defined  in analogy to spin filters based on Mott scattering~\cite{SchoenhenseUM2013} :
\begin{equation}
FoM=  p(\Delta{r}) DE(\Delta{r}).
\label{FoM}
\end{equation}
A numerical simulation for 20~kV incident energy and a ring aperture allowing convergence angles $\alpha \in [8,12]$~mrad is shown in Fig.~\ref{fig:FoM}. A maximum FoM of $3.5 \, 10^{-6}$ is achieved with a detector radius of 0.25~nm. 
It should be noted that the present simplified derivation gives an order of magnitude for standard geometries in the electron microscope. Unconventional setups may result in  dramatically higher values as indicated on the right ordinate of Fig.~\ref{fig:FoM} for 200~V electrons, close to the FoM of Mott detectors. In order to optimize parameters such as convergence angle,  vortex mask and detector design, and thus assess the technical feasibility of such a spin filter,  numerical calculations  including lens aberrations, deviations from the ideal Bessel beam, relativistic effects~\cite{BliokhPRL2011},  the force caused by the gradient of the magnetic field, and the partial coherence of the source are necessary. This is beyond the scope of the present paper and will be discussed elsewhere. 

\section*{Discussion}
The objective lens of an electron microscope leads to spin-to-orbital conversion of an incident unpolarized Bessel beam, and thus can serve as a spin polariser, - even a perfect one in the limit of infinitely small detectors. The effect is  weak in a conventional electron microscope. For standard conditions, the polarisation ("purity")  is of the order of 
$ 10^{-4}$, and the Figure of Merit is $\sim  10^{-5}$. 
As the above estimates show, lower energies increase the performance and the optimum detector size. Decreasing the energy to 200~eV, the FoM comes close to that of conventional Mott detectors which is of the order of $10^{-3}$. 
A key parameter is the convergence angle in the observation plane. One could argue that this angle can be increased by using an immersion lens but it should be noted that a magnetic lens field is compulsory for the STOC effect because electric fields do not rotate the spin. On the other hand, electrostatic immersion lenses could be used to magnify the Bessel beam without deteriorating the STOC effect, thus increasing the useful detector area. 

The dominant obstacle for technical realisation seems to be the small detector size. Apart of suitable magnifying devices one could speculate about the use of nanoclusters or molecules as detectors. In the long run, spin detection of photoelectrons or secondary electrons on the atomic scale appears feasible with this technique.

\section*{Methods}
Here,  derivations of the equations used for the simulations  (done with Wolfram Mathematica~8) are given. 

\subsection*{Spin rotation}


In a constant magnetic field, the spin precesses around the direction of the $B$ field with Larmor frequency $\Omega_L=g|e| B/2 m_e$ where  $m_e$ is the relativistically corrected electron mass  and the electron's gyromagnetic ratio is $g \approx 2$. 
For sufficiently small time $dt$, the magnetic field can be considered constant, and the infinitesimal rotation angle is
$$
d\alpha=\Omega_L \, dt.
$$
According to the Lorentz force, an electron spirals around the direction of the B field with cyclotron frequency $\Omega_C=|e|B/m_e$. Since only the momentum perpendicular to the $B$ field changes, the momentum vector shows also precession around the B field, during time $dt$ the rotation angle is $\Omega_C \, dt$. Since for the electron $\Omega_L=\Omega_C$, the angle between momentum vector and spin vector remains constant. 

More generally, the  operator for spinor rotation over the direction $\vec n$, 
  is
\begin{equation}
R_{{\vec n}}(\alpha)=e^{-i\frac{\alpha}{2} {\vec \sigma}{\vec n}},
\end{equation}
where the Pauli matrices $\sigma$ are the generators~\cite{Sakurai} of the rotation angle $\alpha$.
Eq.~1 
 can be written as
\begin{equation}
R_{{\vec n}}(\alpha)={\mathds{1}} \cos \frac{\alpha}{2}  -i\sin \frac{\alpha}{2} {\vec n}{\vec \sigma}
\label{RotExp}
\end{equation}
with 
\begin{equation}
\vec n \vec \sigma =
\begin{pmatrix}
\cos \theta & -i \sin \theta \, e^{-i\phi'} \\
-i \sin \theta \, e^{i\phi'} &  -\cos \theta 
\end{pmatrix}
\label{msigma}
\end{equation}
where $\theta, \, \phi'$  are the  spherical coordinates of the unit vector $\vec n$ on the Bloch sphere. 

For a cylindrically symmetric lens field, the angle between the rotation axis $\vec n$ for the spinor and the point source is independent of the azimuth $\phi$, so $\phi'=\phi +\phi_0$, and 
Eqs.\ref{RotExp} is
\begin{equation}
R_{{\vec n}}(\alpha)=
 \left( \begin{array}{cc}
\cos \frac{\alpha}{2} -i\sin \frac{\alpha}{2} \cos \theta & -i \sin \frac{\alpha}{2} \sin \theta \, e^{-i(\phi+\phi_0)} \\
-i \sin \frac{\alpha}{2} \sin \theta \, e^{i(\phi+\phi_0)} & \cos \frac{\alpha}{2} + i \sin \frac{\alpha}{2} \cos \theta \end{array} \right)
\label{RotMat}
\end{equation}

\subsection*{Fourier transforms of Bessel beams}
In the co-rotating Larmor system, the wave function in plane B is the Fourier transform of that in plane A:
\begin{eqnarray}
\psi_B(r,\varphi)&=&\tilde \psi_A= \frac{1}{2 \pi}\int_0^{2\pi} d\phi  {\tilde\psi_{Q,\phi}} 
  \, e^{i k z} \nonumber \\
&=&\frac{1}{2 \pi}\int_0^{2\pi} d\phi \, e^{i Q r \cos(\phi-\varphi)}e^{i \phi} e^{i k z} R_{{\vec n}}(\alpha) \, w .
\label{appendix1}
\end{eqnarray}
Extracting the $\phi$-dependent factors from $R_{\vec n}(\alpha)$, we arrive at integrals of the form
$$
\int_0^{2\pi} d\phi \, e^{i Q r \cos(\phi-\varphi)}e^{i \phi}e^{ i l \phi} 
 $$
 with integer $l \in [-1,1]$.
Substituting $\phi'=\phi-\varphi+3 \pi/2$, we have the integral representation of Bessel functions $J_l$,
$$
i^{(l+1)}\,e^{i(l+1)\varphi} \,\int_0^{2\pi} d\phi' \, e^{-i Q r \sin\phi'}e^{i(l+1) \phi'} =i^{(l+1)}\,e^{i(l+1)\varphi} \, 2 \pi \,J_{l+1}(Qr) .
 $$
Arranging terms in Eq.~\ref{appendix1}:
\begin{equation}
\psi_B=
 i \left( \begin{array}{cc}
 J_1(Qr)e^{i\varphi}(\cos \frac{\alpha}{2} -i \sin \frac{\alpha}{2} \cos \theta) & - J_0(Qr) \sin \frac{\alpha}{2} \sin \theta \, e^{-i\phi_0} \\
 J_2(Qr)e^{2i \varphi}\sin \frac{\alpha}{2} \sin \theta \, e^{i\phi_0} &  J_1(Qr)e^{i \varphi}(\cos \frac{\alpha}{2} + i \sin \frac{\alpha}{2} \cos \theta)
 \end{array} \right) w \, e^{ikz}
\label{appendix2}
\end{equation}
from which the transfer matrix $T$
 is immediately obtained. \footnote{Transforming back to the lab system from the corotating Larmor system is simply done by subtracting the angle of image rotation between planes A and B from  $\phi_0$. Since this angle is a free parameter, the structure of Eq.~\ref{appendix2} remains unchanged.}
\begin{equation}
T=
 i e^{i k z}\left( \begin{array}{cc}
 J_1(Qr)e^{i\varphi}(\cos \frac{\alpha}{2} -i \sin \frac{\alpha}{2} \cos \theta) & - J_0(Qr) \sin \frac{\alpha}{2} \sin \theta \, e^{-i\phi_0} \\
 J_2(Qr)e^{2i \varphi}\sin \frac{\alpha}{2} \sin \theta \, e^{i\phi_0} &  J_1(Qr)e^{i \varphi}(\cos \frac{\alpha}{2} + i \sin \frac{\alpha}{2} \cos \theta)
 \end{array} \right)
 \nonumber
\end{equation}

{\bf Acknowledgements:} P.S.  acknowledges the support of the Austrian Science Fund, project I543-N20.

\bibliographystyle{prsty} 



\end{document}